\documentclass[twocolumn ,showpacs,preprintnumbers,amsmath,amssymb,prl, superscriptaddress]{revtex4}

\usepackage{graphicx}
\usepackage{dcolumn}
\usepackage{bm}
\usepackage{color,ulem}

\begin{document}

\title{Partial nonlinear reciprocity breaking through ultrafast dynamics in a random photonic medium.}
\author{Otto L. Muskens}
\affiliation{School of Physics and Astronomy, University of
Southampton, Highfield, Southampton SO17, 1BJ, United Kingdom}\email{O.Muskens@soton.ac.uk}

\author{Paul Venn}
\affiliation{School of Physics and Astronomy, University of
Southampton, Highfield, Southampton SO17, 1BJ, United
Kingdom}

\author{Timmo van der Beek}
\affiliation{FOM Institute AMOLF, Science Park 104, 1098 XG Amsterdam, The Netherlands}

\author{Thomas Wellens}
\affiliation{Physikalisches Institut der Albert-Ludwigs Universität Freiburg, Hermann-Herder Str. 3, D-79104 Freiburg, Germany}

\begin{abstract}
We demonstrate that ultrafast nonlinear dynamics gives rise to
reciprocity breaking in a random photonic medium. Reciprocity
breaking is observed via the suppression of coherent
backscattering, a manifestation of weak localization of light. The
effect is observed in a pump-probe configuration where the pump
induces an ultrafast step-change of the refractive index during
the dwell time of the probe light in the material. The dynamical
suppression of coherent backscattering is reproduced well by a
multiple scattering Monte Carlo simulation. Ultrafast reciprocity
breaking provides a distinct mechanism in nonlinear optical media
which opens up avenues for the active manipulation of mesoscopic
transport, random lasers, and photon localization.
\end{abstract}
 \maketitle

There is a strong analogy between the transport of classical waves
such as light and sound in scattering media, and the mesoscopic
physics of quantum waves, such as electrons in a solid, or matter
waves in an optical speckle potential \cite{Lagendijk2009}.
Whereas in many quantum systems coherence is affected by inelastic
processes, interactions are generally weak for light and coherent
effects can be studied in large-scale systems. The role of
interactions and decoherence in many-body quantum systems can be
simulated by introducing a nonlinear optical response
\cite{Skipetrov01, Spivak00, Conti07,
Shadrivov10,Conti11,Wellens09}. Recent pioneering
experiments have explored nonlinear light scattering in Kerr-media
and cold atomic clouds in the stationary regime
\cite{Schwarz07,Sebbah11,Labeyrie08,Wilkowski05}.

In this Letter, we demonstrate partial breaking of the
reciprocity symmetry in a random scattering medium on ultrafast
time scales through the dynamics induced by a high-intensity,
femtosecond pump pulse. As a probe for the breaking of reciprocity
we use the coherent backscattering effect, a manifestation of weak
localization for optical waves. Coherent backscattering (CBS) is
the constructive interference of reciprocal light paths, resulting
in a cone of enhanced intensity around the backscattering
direction \cite{Albada85}. The center of the cone results from the
constructive interference of very long light paths, which makes
CBS a sensitive probe for multiple scattering wave transport.

The reciprocity breaking effect is induced through ultrafast
dephasing of light paths occurring in strongly scattering
nanostructured semiconductors \cite{Abb11}. Reciprocity breaking
requires the dynamics to be fast on the time scale of the photon
dwell time in order to achieve an asymmetry between the direct and
reciprocal paths. Such a regime of \textit{adiabatic} control over
light has been achieved only recently in high-quality photonic
crystal nanocavities \cite{Notomi2010,Ctistis2011}, and is of
interest for applications in controlled storage and release of
optical information. Optical devices exploiting concepts from
mesoscopic transport are currently receiving increasing interest
\cite{Putt2011,Sap2010,Ler2007,Wiersma2008}. A degree of active
control over the flow of light in these devices would be of
interest for applications.

\begin{figure}[t]
\centering
\includegraphics[width=8.0cm]{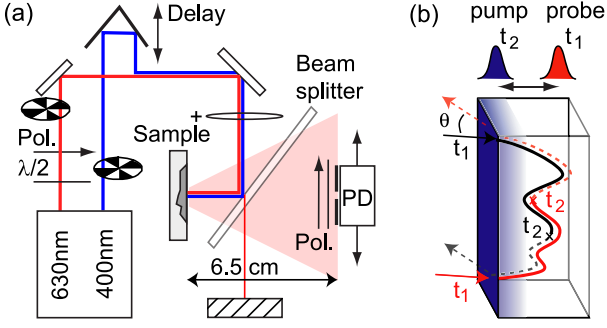}
\caption{\label{fig:setup} (a) Experimental setup for
time-resolved pump-probe coherent backscattering spectroscopy. (b)
Principle of reciprocity breaking, where the symmetry of
reciprocal multiple scattering path (arrows) is changed by a
delayed pump pulse ($t_2>t_1$).}
\end{figure}

The reciprocity breaking was measured in a slab of porous
GaP of $81 \pm 2$~$\mu$m thickness, obtained by etching of a GaP
wafer. Pump-probe nonlinear CBS cones were measured in a
beamsplitter configuration as shown in Fig.~\ref{fig:setup}(a).
The frequency-doubled output from a regeneratively amplified
Ti:Sapphire laser (Coherent RegA, $200 \pm 10$~fs , 250 kHz) at
400 nm wavelength was used as a pump for inducing an ultrafast
nonlinear response. As a probe the signal output from a parametric
amplifier at 630 nm wavelength was used. Both beams were focused
to a 25-$\mu$m diameter spot onto the sample using a lens with a
focal length of 15~cm to achieve a sufficient pump fluence for
nonlinear modulation, resulting in an angular resolution of
20 mrad. Backscattered light was collected using a Si-photodiode
with a 0.5-mm aperture, mounted on a translation stage positioned
6.5~cm away from the sample. Linear polarization filters were used
to select either the polarization conserving ($R_{//}$) or the
nonconserving ($R_{\perp}$) scattering channels. The diffuse and
CBS intensities $R_{\perp}$ and $R_{//}$, as well as the
differential pump-probe signals
 $\Delta R_{\perp}$ and $\Delta R_{//}$ , i.e. the difference of the probe signal in presence of the pump with respect to $R_{\perp}$ or $R_{//}$, respectively, were measured using
lock-in detection. All signals were normalized to the diffuse reflectance $R_\perp$ to divide out the angle-dependence of the setup and the (Lambertian) diffuse intensity distribution.

For the interpretation of our experimental results, Monte Carlo
(MC) numerical simulations of the transient response were produced
by spatiotemporal tracking of photon trajectories in a finite
slab. Monte Carlo is a technique suitable for simulating diffuse
transport and coherent backscattering of light \cite{Xu08}. In the
current model, illustrated in Fig.~\ref{fig:setup}(b), the pump
pulse induces a change in the complex refractive index at a time
$t_2$. In all simulations in this Letter, we used one set of
optimized parameters. These include the a-priory known
parameters of sample thickness, laser pulse width and instrumental
resolution. An internal reflection probability of 0.6,
corresponding to a linear refractive index $n_0=1.47$, was
included, following measurements on similar porous GaP slabs
\cite{Rivas03}. Parameters extracted from comparison between
numerical and experimental data (shown below) include the (linear)
transport mean free path $\ell_e=0.45 \pm 0.1$~$\mu$m
(corresponding to $k_0\ell_e=4.5$, with $k_0=2\pi/630$~nm$^{-1}$
the wavevector of the probe light in vacuum) and the (linear)
transport mean free time $t_e=5 \pm 1$~fs, the spatiotemporal
refractive index $n(z,t)=n_0+\Delta n(z,t)$, where the nonlinear
index change $\Delta n(z,t_2)=(1.4\times 10^{-2}+ 1.4\times
10^{-3} i)\exp(-z/L_{\rm exc})$ is contained within a
photo-excited surface region of width $L_{\rm exc}=0.2$~$\mu$m
just after absorption of the pump pulse at time $t_2$. The
imaginary part of $\Delta n$ describes photo-induced absorption of
the probe light, with absorption time
$\tau_{\rm abs}=n_0/(2 ck_0 {\rm Im}\Delta n)=0.17$~ps
at $z=0$
and time $t_2$. Subsequent spatiotemporal dynamics involve a
rapid, picosecond expansion of the excited volume due to carrier
migration, with a concomitant decrease in the nonlinear refractive
index change $\Delta n$, such that the total amount of excitation
in the volume remains unchanged by this expansion. In addition,
the excitation is assumed to decay with a time constant of 30~ps,
which, however, hardly affects the results presented in this
Letter.

\begin{figure}[t]
\centering
\includegraphics[width=8.7cm]{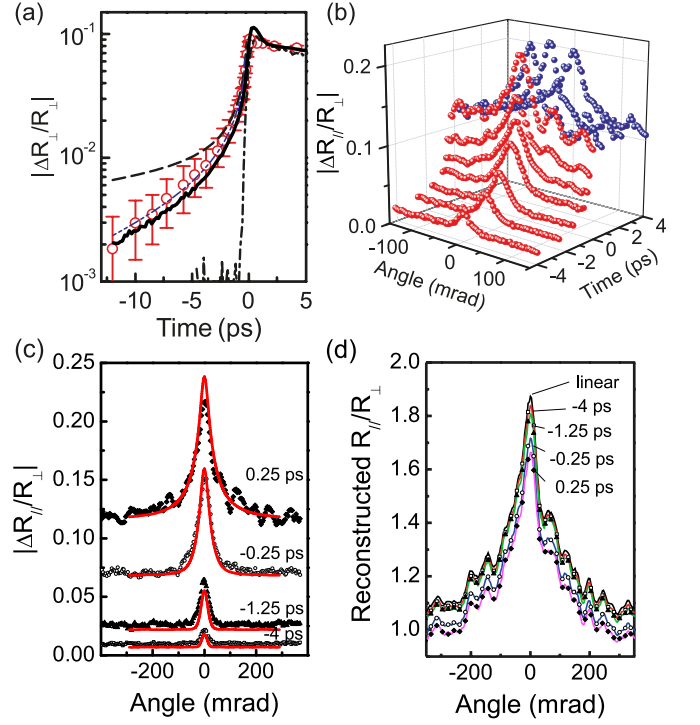}
\caption{\label{fig:3dcones} (a) Differential diffuse reflectivity
$\Delta R_\perp/R_\perp$ as a function of the time delay between
pump and probe pulse. Lines indicate model calculations $\propto
t^{-1/2}$ (dash), same but including an exponential cutoff with
time-constant $10.2\pm 0.7$~ps (dotted, blue), and Monte Carlo
model (thick line). Dash-dotted line: signal from a homogeneous
(i.e. non-disordered) GaP substrate. (b) Angle- and time-resolved
differential reflectivity $\Delta R_{//}/R_\perp$ in the
polarization conserving channel, for negative (red) and positive
(blue) time delays $t_1-t_2$. (c) Same as (b) for several delay
times, including results from MC model (red lines). (d) CBS
intensity traces without pump ('linear') and with pump
reconstructed using the differential response curves from (c).}
\end{figure}

Figure~\ref{fig:3dcones}(a) shows time-resolved
reflectivity changes $\Delta R_\perp/R_\perp$ obtained in the
cross-polarized, diffuse scattering channel. The signal at
positive pump-probe time delays is characteristic for the excited
semiconductor material and resembles our earlier measurements for
bulk GaP (dash-dotted line) \cite{Abb11}. At negative delay times,
an additional tail is present for the scattering layer. This tail
corresponds to the contribution from photons entering the sample
before the pump and retained inside the sample for thousands of
scattering events before undergoing photo-induced absorption
initiated by the pump pulse.

The negative-time tail of Fig.~\ref{fig:3dcones}(a) thus acts as a
probe of the distribution of photon dwell times in the scattering
medium, complementary to time-resolved interferometry experiments
\cite{JohnsonPRB03,Toninelli08}. Generally, the pump-probe signal
can be expressed as a convolution of the diffuse path-length
distribution $\alpha_d(t)$ and a photo-induced absorption term
$F(t)$, yielding
\begin{equation}\label{eq:pp}
\frac{\Delta
R_\perp(t)}{R_\perp}=\frac{1}{\alpha^{\rm tot}_{d}}\int_{0}^\infty dt_2
\alpha_d(t_2- t) F(t_2) \, ,
\end{equation}
where $t=t_1-t_2$ is the pump-probe delay time. The total albedo $\alpha^{\rm tot}_{d}$ results from integrating
$\alpha_d(t)$ over time. In the diffusive limit, the path length
distribution is given by $\alpha_d(t)\propto t^{-3/2}$ for times
larger than the extinction mean free time $t_e$
\cite{AkkermansBook}. For the simple case of diffusion and a step
response $F(t)=\eta_0 \Theta(t)$, Eq.~(\ref{eq:pp}) can be solved
analytically, yielding $\Delta R_\perp(t)/R_\perp = \eta_0
|t/t_e|^{-1/2}$ for $t<-t_e$. This relation is shown by the dashed
line in Fig.~\ref{fig:3dcones}(a). Clearly, this curve
overestimates the contribution of long paths in the tail for
$t<-2$~ps. This discrepancy can be corrected by including an exponential cutoff in the
distribution $\alpha_d(t)$ with
time constant of $10.2 \pm 0.7$~ps corresponding to $\sim 2\times 10^3$ scattering events [dotted line
(blue) in Fig.~\ref{fig:3dcones}(a)]. The loss of long light paths in the pump-probe signal is related to the finite system size as explained below. Our measurements show that pump-probe transient absorption forms a sensitive probe of very long light paths in the photonic medium.

The above form of $F(t)$ is valid under the assumption that only a
thin surface layer is affected by the pump pulse, which is a good
approximation given the bulk absorption length of GaP of around
0.2~$\mu$m at 400~nm wavelength \cite{Palik}. Therefore,
the relative amount of absorption is approximately constant
($\eta_0\simeq 10^{-1}$) for all probe photons which undergo a
scattering sequence whose length exceeds the pump-probe delay time
$t$. This picture is confirmed by the MC model (thick
solid line). It was found from the MC model that the slope of the
negative-time tail is determined entirely by the 25-$\mu$m transverse spot size of the pump beam. Long light paths diffuse out of the excited region, resulting in the observed reduction of the long-time tail. The spatiotemporal expansion mainly affects the
decay at positive times in Fig.~\ref{fig:3dcones}(a). The real
part of $\Delta n$ does not affect the diffuse tail; however it is
mainly responsible for the reciprocity breaking effect as shown
below.

\begin{figure}[t]
\centering
\includegraphics[width=8.6cm]{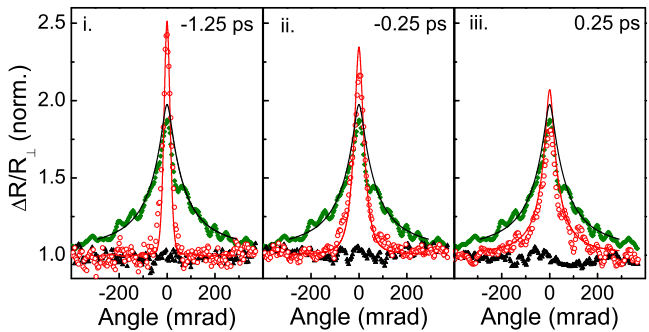}
\caption{\label{fig:traces} Differential reflectivity for parallel
(open circles, red) and cross-polarized (closed triangles, black)
channels normalized to their values at $\theta=-270$~mrad. Closed
diamonds, green: Linear coherent backscattering cones. Lines:
results from Monte Carlo model for linear and differential
reflectivity cones. For negative delay times, the
differential reflectivity in the parallel channel is larger than
2, which indicates the occurence of partial reciprocity
breaking.}
\end{figure}

We proceed to investigate the effect of ultrafast excitation on
the CBS intensity in the polarization conserving channel.
Figure~\ref{fig:3dcones}(b) shows the experimental differential
reflectivity $\Delta R_{//}/R_\perp$. We emphasize that
these nonlinear CBS traces are obtained as the difference between
two CBS intensity cones respectively in presence and absence of
the pump pulse, as is typical for lock-in detection. Therefore,
the differential reflectivity corresponds to the intensity which
is \textit{removed} from the linear CBS cone as a consequence of
photo-induced effects. This can be observed in
Fig.~\ref{fig:3dcones}(c,d) where we have used the differential
signals to reconstruct the modified intensity CBS cones at various
delay times. The cone rounds off and effectively broadens as long light paths are removed. Both the coherent backscattering and the diffuse
background show a maximum decrease of around 12\% around the
pump-probe delay of 0~ps.

\begin{figure}[t]
\centering
\includegraphics[width=7.8cm]{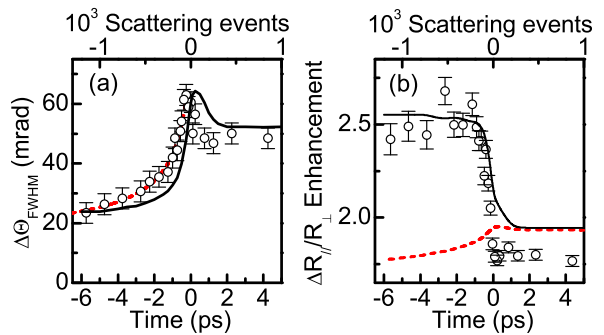}
\caption{\label{fig:theta} (a) Full width at half maximum values
of the CBS cone width, with calculation using the CBS albedo
Eq.~(\ref{eq:ac}) (dashed line, red) and Monte Carlo model (thick
line). (b) Experimental pump-probe enhancement factors (circles)
in comparison with the Monte Carlo model for the full, complex
nonlinear refractive index as described in text (line) and using
only the imaginary part of $\Delta n$ (dashed line, red). }
\end{figure}

The contribution of reciprocity breaking to the overall
pump-probe signal can be analyzed by comparing the ratio of of the
intensities removed from the CBS with those removed from the
diffuse albedo. In particular, the more the total CBS cone
\textit{decreases} due to reciprocity breaking, the more the
differential CBS cone \textit{increases}, which may thus assume a
relative height of larger than 2. Figure~\ref{fig:traces} shows
the three differential CBS cones i, ii, iii of
Fig.~\ref{fig:3dcones}(c) (dots, red), where the change in the diffuse background at -270~mrad is normalized to 1. For comparison, also the diffuse pump-probe
signals are shown (triangles, black). In comparison to the linear
CBS intensity cone (diamonds, green), the differential pump-probe
cones are narrower and have a higher amplitude. The values for the
angular width and differential cone height are shown in
Fig.~\ref{fig:theta}(a,b) against pump-probe delay time.

To understand the trends in Fig.~\ref{fig:traces} and
\ref{fig:theta}, we start from the characteristic time- and
angle-dependent albedo of the CBS cone given by
\cite{AkkermansBook}
\begin{equation} \label{eq:ac}
\alpha_c(\theta,t) \simeq \alpha_d(t)e^{-\frac{1}{3}(k_0 \ell_e
\theta)^2t/t_e} \, ,
\end{equation}
with $\theta$ the backscattering angle. At exact backscattering,
Eq.~(\ref{eq:ac}) reduces to the diffuse albedo,
$\alpha_c(0,t)=\alpha_d(t)$. The angular range $\theta$ over which
paths of a characteristic time duration $\tau_C$ contribute to the
CBS-cone is given by $\theta(\tau_C)=( 3 t_e/k_0^2 \ell_e^2
\tau_C)^{1/2}$. This relation shows that the angular width is
reduced proportional to the square root of the path time duration.
At long negative delay times $t=t_1-t_2$, only long light
paths with dwell times exceeding $|t|$ contribute to the nonlinear
signal, and hence $\tau_C=|t|$. This relationship --
which clearly explains the fact that the nonlinear CBS
traces are narrower than the linear ones, as already observed in
Fig.~\ref{fig:traces} -- is indicated by the dashed line (red) in
Fig.~\ref{fig:theta}(a). The calculated curve includes the
instrumental resolution of 20~mrad, and was fitted to the
experimental data at negative delay times in order to determine
$t_e$. Around $t=0$, the cone width exhibits a maximum, which is
reproduced by the MC simulation (thick solid line) and can be
attributed to the spatio-temporal dynamics of the excited surface
region after excitation. In the MC simulation, the latter is
modelled as an expansion of the excited surface from $L_{\rm
exc}=0.2$~$\mu$m to $2.25$~$\mu$m within a time of $2$~ps. Due to
this expansion, the size of the excited region encountered by the
probe photons increases for positive delay times, leading to
stronger absorption of long scattering paths, and consequently, to
a narrower differential cone.

The maximum height of the $\Delta
R_{//}/\Delta R_\perp$ enhancement factor is shown against delay time in
Fig.~\ref{fig:theta}(b). If this factor is larger than 2, this
indicates that relatively more intensity is removed from the CBS-
than from the diffuse scattering-albedo. Figure~\ref{fig:theta}(b)
shows that this additional contribution is only present at
negative delay times, i.e. for probe photons arriving
\textit{before} the pump pulse. This behavior is reproduced by our
MC simulations, and can be attributed to the asymmetry between
direct and reciprocal light paths induced by the nonlinear
refractive index change \cite{remark3}. Only changes fast with respect to the
photon dwell time are able to break reciprocity. Such fast changes
are mainly obtained at the leading edge of the ultrafast nonlinear
response of GaP. As long as the probe photons arrive before this
nonlinear edge, a sufficient reciprocity breaking is obtained to
produce an effect in the nonlinear CBS-enhancement. Given
the maximum value of 2.5 for the $\Delta R_{//}/\Delta R_\perp$
ratio and our observation that the diffuse scattering albedo of
long light paths is suppressed by $\eta_0\simeq 10\%$, we conclude
that the CBS albedo of these light paths is reduced, on average,
by approximately $15\%$, corresponding to a partial reciprocity
breaking of about $5\%$.

Dephasing and absorption both result in an asymmetric reduction of
the amplitudes, and thus a reciprocity breaking effect. In the MC
simulation (line) in Fig.~\ref{fig:theta}(b), good agreement with
experiments is found when assuming for the real part to be ten
times larger than the imaginary part of $\Delta n$. This implies
that the effect of dephasing is much stronger, consistent with our
earlier results \cite{Abb11}. A purely imaginary $\Delta n$
(dashed line) provides only a small reciprocity breaking effect
which cannot explain
 the experimental enhancement factors.

In conclusion, we have demonstrated partial suppression of
reciprocity in a random medium through the coherent backscattering
effect. Reciprocity is a fundamental property of photonic media,
and its ultrafast control opens up a new approach to nonlinear
manipulation of photonic eigenstates. Our work is the first step
toward the control of mesoscopic interference phenomena in random
media. We achieve 5\% reciprocity breaking using only a 200-nm
thin excitation region. Typically, localization lengths in random media and gain lengths in random lasers amount to at least several micrometers \cite{Lagendijk2009, Wiersma2008}. Ultrafast reciprocity breaking of closed loops and random laser modes is therefore feasible but requires further optimization of the pumping conditions to achieve dephasing in a large excitation volume. This may be achieved using a longer pump absorption length, two-photon absorption, or use of two-dimensional waveguides where pump light can be coupled in from the top of the structure.


We thank the FASTlab and D. Kundys for support, and A. Lagendijk
for stimulating discussions. This work was supported by EPSRC
through grant EP/J016918/1 and by DFG through grant BU
1337/8-1.



\begin{thebibliography}{10}
\bibitem{remark3} The difference in linear and nonlinear CBS enhancement factors between MC simulations and experiments results from a reduced experimental CBS enhancement factor for the linear cone which was not included in the model.
\end{thebibliography}
\end{document}